\documentclass[9pt,twocolumn,twoside]{opticajnl}
\journal{opticajournal} 

\setboolean{shortarticle}{true}

\usepackage{lineno}
\usepackage{sidecap}
\usepackage{authblk}
\usepackage{pdfpages}

\title{Broadband and wide-angle beam deflection enabled by dynamically reconfigurable meta-arrays}

\author{Koffi-Emmanuel Sadzi and Abdoulaye Ndao\textsuperscript{*} \\

{\small University of California, San Diego, Department of Electrical and Computer Engineering, La Jolla, California, United States\\

\textsuperscript{*}a1ndao@ucsd.edu}}

\begin{abstract}
We present a structurally simple yet functionally versatile reflective meta-array composed of phase-change Antimony trisulfide ($\rm Sb_2S_3$)   nanorods enabling broadband and wide-angle beam deflection in near-infrared. The device achieves over 80\% deflection efficiency over a 1000 nm wide passband, and covering from O-band (1260 nm-1360 nm) to U-band (1565 nm-1625 nm) in the \emph{amorphous} state. Meanwhile, in its \emph{crystalline} state, we see a reduction of the efficiency to 40\% on average with a maximum passband of 800 nm over the C-band. Moreover, its simple architecture simultaneously enables spectral filtering and beam splitting, delivering a compact, multifunctional solution optimized for high power applications.

\end{abstract}

\setboolean{displaycopyright}{false} 

\begin{document}

\maketitle

\section{Introduction}

Controlling the propagation of light is fundamental to optical science and technology. In that regard, an increase in interest in wavefront engineering has led to diverse approaches \cite{Wood2935,Sergei2016,Liu2022}, away from the traditional methods relying on bulky components like lenses and prisms, which are shaped according to Snell's law. Two of such approaches, anomalous reflection and its counterpart, anomalous refraction, have become a fundamental tool in modern optics, finding applications in optical communications \cite{Rose2005}, laser systems \cite{Strickland1985}, and pulse compression technologies \cite{Hawthorn2001, Perry1994, Rhee1994}. Anomalous reflection, often referred to as deflection, describes the unusual steering of incident electromagnetic waves with a non-specular angle, namely, towards a direction not predicted by Snell's law. Deflection typically relies on directing most of the incident wave energy into a non-zero diffraction order by design. Therefore, high deflection efficiency, combined with wide angular and broadband operation, is highly desirable.

To satisfy the stringent demands of modern optical systems, a wide range of grating-based strategies have been investigated. These include binary blazed gratings \cite{Lee2000,lalanne1999}, multi-ridge complex gratings \cite{Hehl1999}, and structures incorporating artificial materials \cite{Sauvan2004}—particularly effective in the near-infrared and visible spectral regions. Alternative approaches have employed photonic crystal gratings (PhCGs) \cite{Gralak2000,Serebryannikov2009}, which exploit Bloch mode dispersion for tailored light manipulation. 

Recently, ultrathin optical interfaces composed of subwavelength nanostructures named metasurfaces \cite{lalanne_optlett_1998,bomzon_optlett_2001,lu_opexpress_2010,fong_antennaspropag_2010,yu_science_2011,ndao_natcomm_2020,Hsu2017,Ndao2020exosome} have rapidly emerged as a disruptive technology in flat optics. By offering unprecedented control over the phase, amplitude, and polarization of light at the nanoscale \cite{Genevet_Capasso2017,Hsu2017FromParabolicToMetasurface,Lee2018,Yang2021Tunable,Yang2022Plasmonic}, they enable functionalities far beyond those of conventional bulk optics. This paradigm shift has led to the demonstration of a broad spectrum of devices, including metasurface deflectors \cite{Yu2015, M_Shalaev2015,Ha2018PlanarLens,Ndao2018Plasmonless}.

In the context of high-power applications, a desired additional functionality is variable efficiency or power-limited features. Few existing designs \cite{Vella2016, Valentine18, Capasso2017, Valentine2020, Hsu_Nado2021, Parra2021, Choi2024} offer built-in power-limiting capabilities; however, they do not integrate broadband, wide-angle, and spatial filtering.

In this work, we propose and numerically demonstrate a simple periodic meta-array composed of $\rm Sb_2S_3$ nanorods, which enables wide-angle, wideband anomalous reflection. The design naturally incorporates additional features, such as spatial filtering and beam splitting, without requiring extra tuning parameters.{\color{black} The same device can be reconfigured for high power applications exploiting the absorption of the crystalline-$\text{Sb}_2\text{S}_3$}. This approach opens new avenues for the realization of compact, high-power-compatible optical elements, well-suited for integrated photonics and sophisticated wavefront control.

\section{Results and Discussion} 

\begin{figure}[!ht]
 \centering
\includegraphics[width=\linewidth]{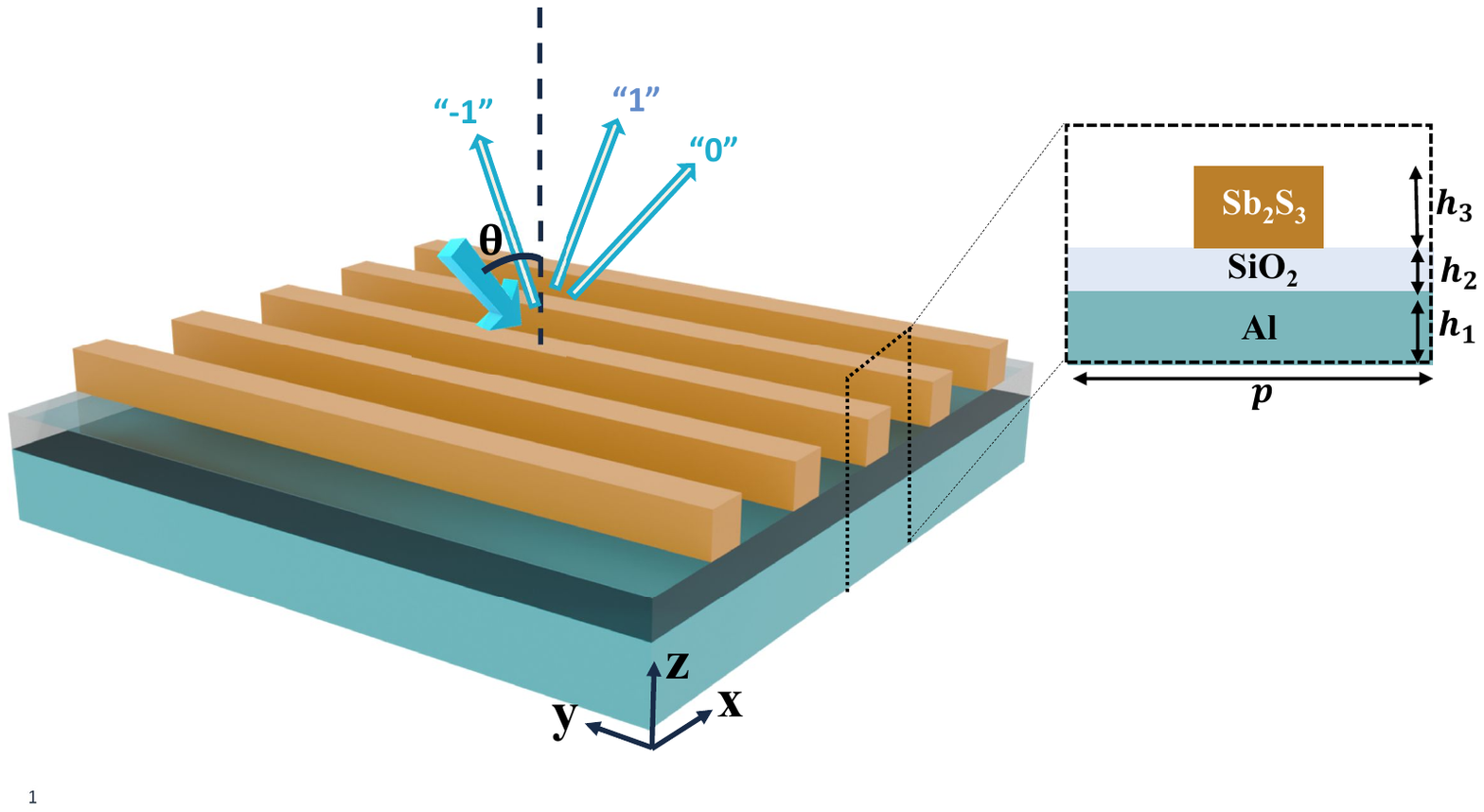}
 \vspace{-1cm} 
\caption{ Schematic of the  proposed device: $\theta$ represents the incident angle; the reflected beam includes the 0th and -1st diffraction orders, along with other deflected components such as 1st order illustrated here. (\textit{right}) Cross-section of a unit cell of the meta-array: $p=1150\text{ nm} $ represents the period of the unit cell, $w=300\text{ nm}$ denotes the width of the nanorod, and $h_1=100\text{ nm},h_2=60\text{ nm},h_3=200\text{ nm}$ represent the heights of the Al layer, of the $\rm SiO_2$ layer and the $\rm Sb_2S_3$ nanorod, respectively}
\label{3d_schem}
\end{figure}
Our reflective meta-array consists of rectangular nanorods made of Antimony trisulfide $\rm Sb_2S_3$, a phase-change material (PCM) \cite{Bieganski2024}. The PCM has two phases: \emph{crystalline}, which is lossy ({\color{black} see Supplement 1 for details}), and low-loss \emph{amorphous} in the bandwidth of interest (O-U band). {\color{black} This prominent material enables the designed device to achieve an ultra-wideband and wide-angle wave deflections as well as a configurability to accommodate high powers applications. Namely, under high power continuous-wave or pulsed lasers, the switching property of the PCM allows a passive transition from the amorphous to crystalline state for a continuous-wave source when the temperature in the PCM reaches its transition temperature (refer to Supplement 1 for details on switching mechanism and threshold) }. In principle, the incident wave is reflected using the $n=-1$ diffraction order in the case of transverse electric (TE) polarization, where the electric field vector of the incident wave is parallel to the axes of the nanorods.

Figure~\ref{3d_schem} represents the schematic of our proposed structure, as well a cross-section.  The structure contains a periodic array of rectangular $\rm Sb_2S_3$ nanorods, patterned on the top of $\rm SiO_2$ buffer layer deposited on the Aluminum  ($\rm Al$) layer.  $p$, $w$ denote the period of the unit cell, the width of each nanorod, respectively; $ h_1, h_2,h_3$ are the thicknesses of the Aluminum layer, of the Silicon Oxide layer, and the PCM nanorod, respectively. To operate in reflection mode, we use $h_1=100$ nm and while optimizing for the difference in the efficiency of deflection at \emph{amorphous} and \emph{crystalline} state, as well as a steep cutoff near O- and U- band, the designed geometrical parameters are: $p=1150\,\text{ nm},w=300\text{ nm},h_1=100\text{ nm},h_2=60\text{ nm},h_3=200\text{ nm}$. Those values are selected to ensure robustness of the design, anticipating even significant fabrication imperfections.

Owing to the fundamental impact of loss on system performance, the efficiency of the $n$th order deflection ($\eta_{(n)}$) of the device is evaluated as a ratio of the power converted to the $n$th order to the total incident power. Considering an incident wave, with wavelength $\lambda$, on a grating structure of period $p$ and at an incidence angle $ \theta$, the angle $\phi_n$ of the reflected waves following the diffraction orders ($n$) are given by the well-known relation \cite{Petit1980}:

\begin{align}
    \sin \phi_n =\sin \theta +n\,\lambda /p
    \label{diff_rel}
\end{align}
A given order $n$ is reflected or propagates when $|\sin \phi_n|<1$. This supports that the 0th order is always reflected as a specular reflection. As achieved and shown below, when the structure support only $-1$ order as nonzero propagating mode, a high efficiency is observed both for the \emph{crystalline} and \emph{amorphous} phase of the nanorod. 
\begin{figure}[ht]
 \centering 
 \includegraphics[width=\linewidth]{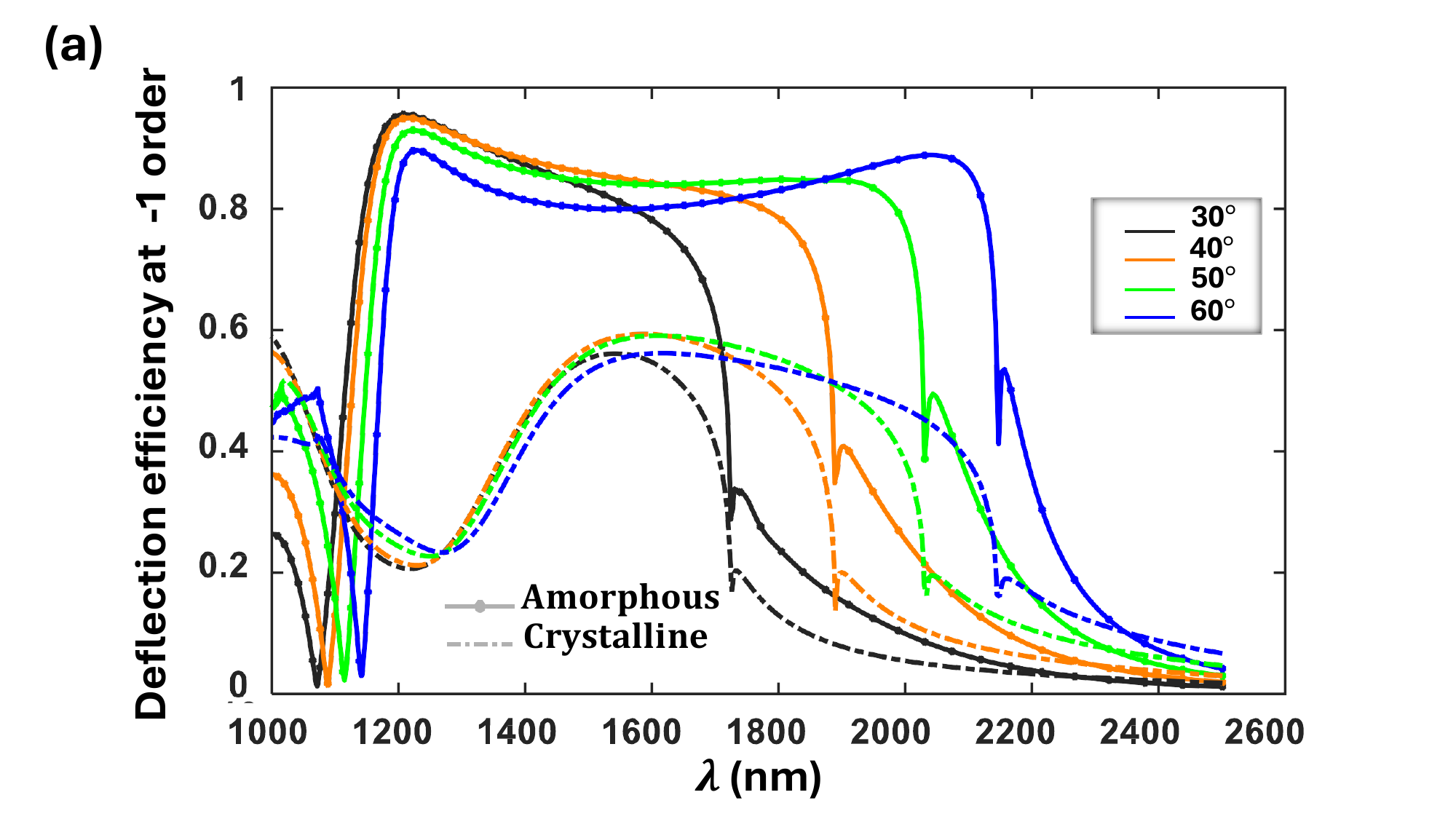}
 \hfill
 \includegraphics[width=\linewidth]{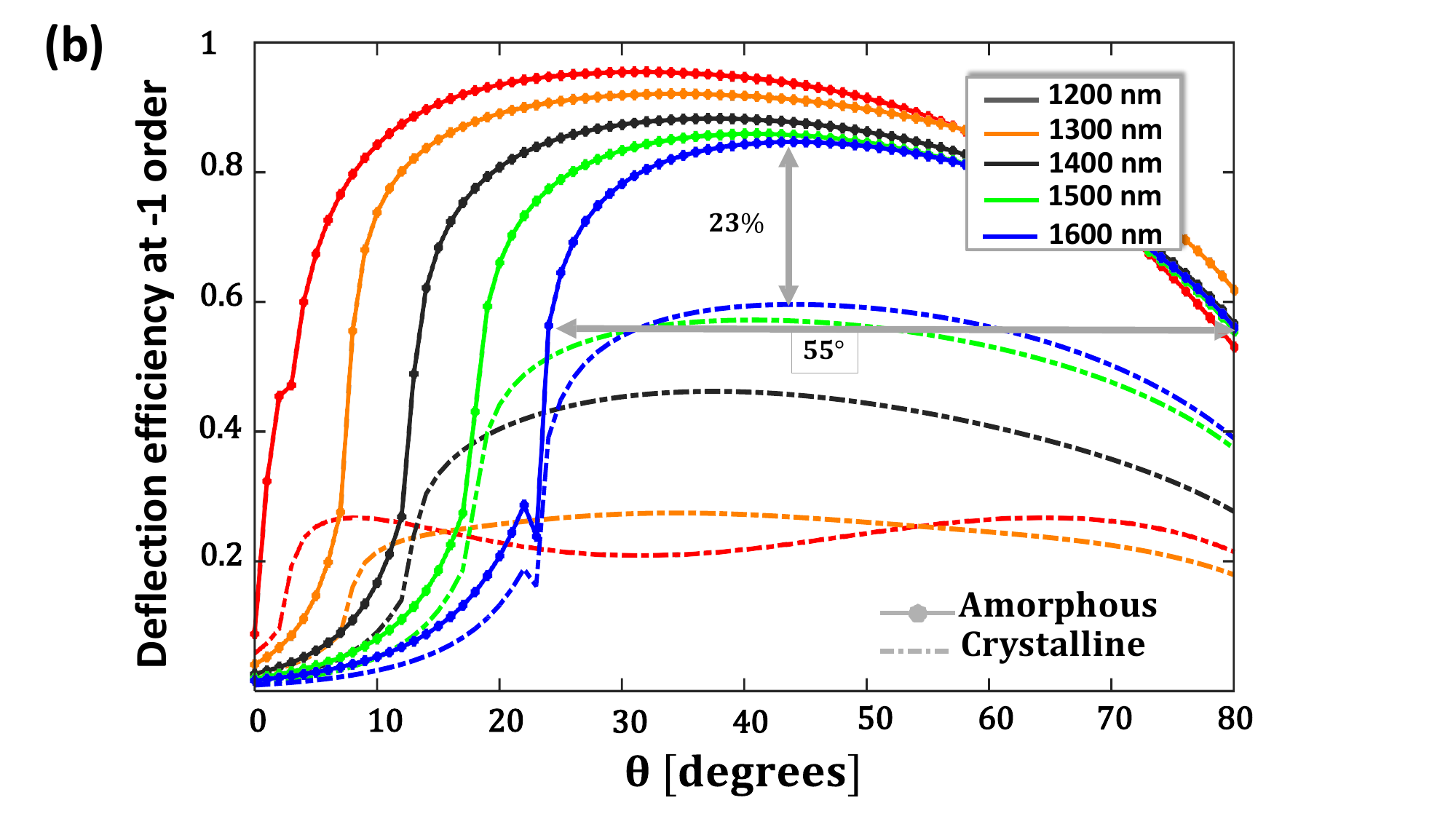}
\caption{ Simulation results of the -1st order deflection efficiency, as a function of (a) the wavelength $\lambda$ and (b) the incidence angle $\theta$ for both \emph{amorphous} (in solid line with marker) and \emph{crystalline}  (dot-dash line) phases. TE polarization is considered. The wideband deflection achieved covers O-band to U-band in the case of the \emph{amorphous} state (solid line) with efficiency greater than 80\%. In the \emph{crystalline} phase case, a deflection efficiency as low as 20\% can be achieved (in O-band). (b) shows that despite the reduction of angle range with wavelength, a minimum of $55^\circ$ is achieved with a difference of 23\% between the \emph{crystalline} and \emph{amorphous} state.}
\label{deflect_meas}
\end{figure}

To examine the performance of our design, we perform a full-wave simulation using   Computer Simulation Technology (CST) software. Figure~\ref{deflect_meas}(a) compares, under TE polarization incidence wave, the deflection efficiency spectrum at the $-1$st order for the PCM in \emph{crystalline} and \emph{amorphous} states. In the \emph{amorphous}, a passband of over 1000 nm wide can be achieved, with very sharp cutoffs at both short and long wavelengths. An efficiency of over $80\%$ also characterizes the passband. Hence, this case shows a good performance as a spectral filter. In contrast, the \emph{crystalline} case exhibits a gradual roll-off and can still achieve a passband of 800 nm, covering the entire U-band. It yields, as expected, a low efficiency, limiting the reflected power below $60\%$ and as low as $\sim 20\%$. Hence, this case demonstrates{ \color {black}the versatility of the device for power sensitive applications}. 
\begin{figure}[!h]
 \centering
\centering \includegraphics[width=\linewidth]{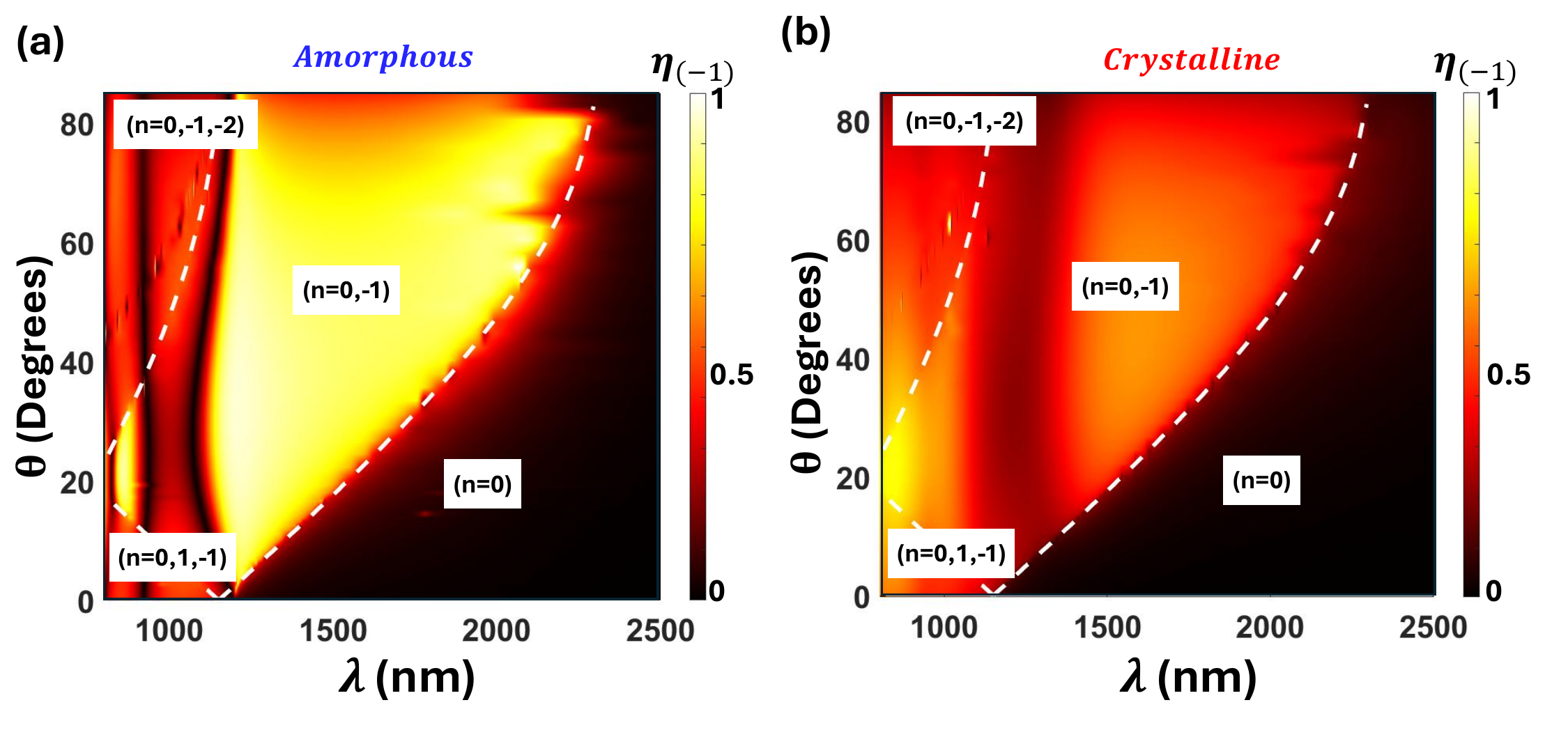}
\hfill
\includegraphics[width=\linewidth]{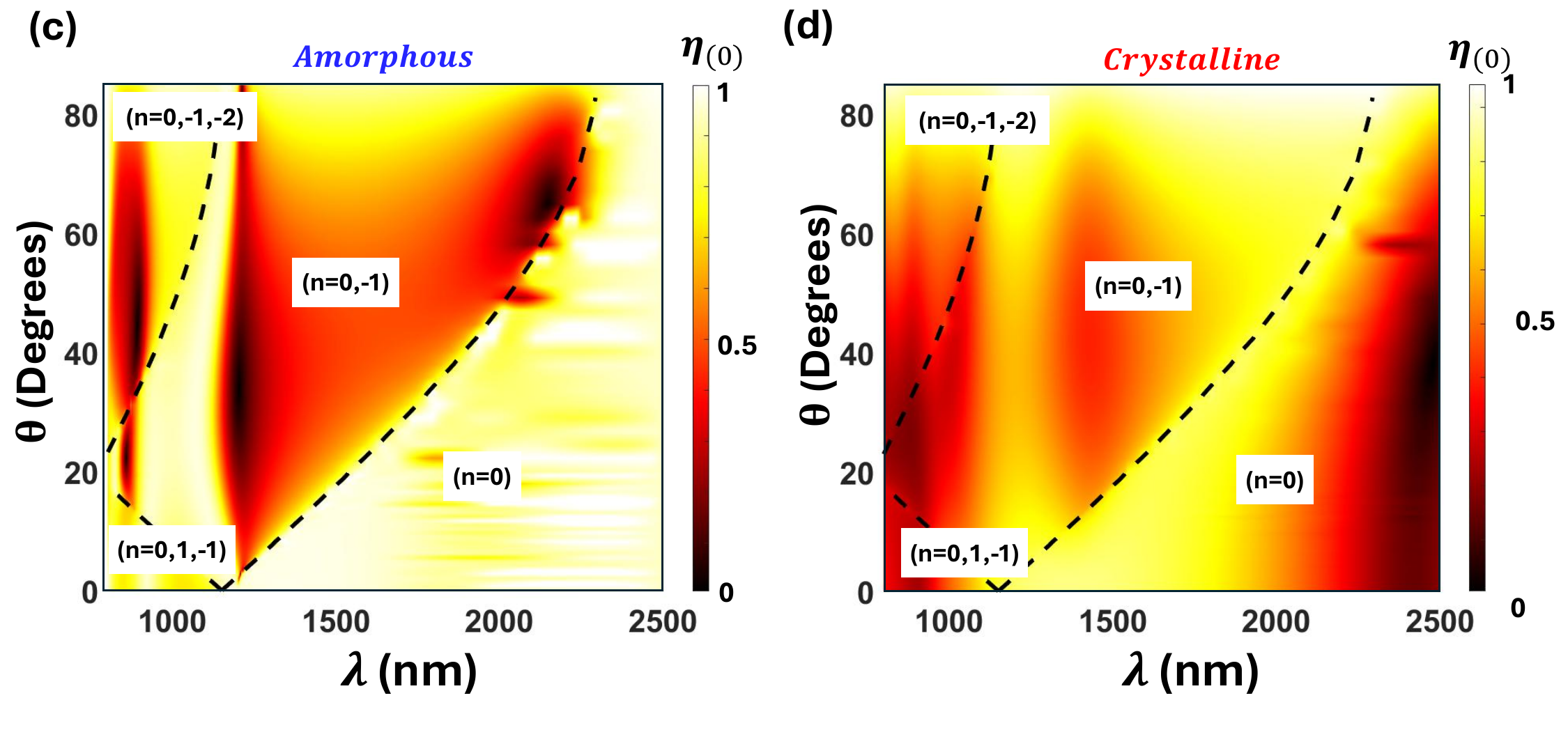}
\caption{Contour plot of the deflection efficiency, $\eta_{(n)}$, for $n= -1$ (a-b) and  $n = 0$ (c-d) order as a function of both wavelength and the incidence angle. The PCM is  \emph{amorphous} in (a,c) and \emph{crystalline} in (b,d) simulations. The region to the left of the dashed black lines constitutes the domain where the corresponding order ($n = -1, -2 \text{ or} +1$) propagates. }
\label{contours}
\end{figure}

Moreover, to confirm the wide-angle deflection of the design, we analyze the efficiency versus the $\theta$ for fixed wavelengths. As shown in Fig.~\ref{deflect_meas}(b), the \emph{amorphous} case maintained an efficiency $\eta_{(-1)}$ over a wide range of $\theta$, spanning over $50^\circ$ with the capability of a slight angle for short wavelengths. 
It enables also an angular passband with a sharp low-$\theta$ cutoff. Interestingly, the \emph{crystalline} case maintains a low and nearly constant efficiency.
One can also notice that around $45^\circ$, the difference in efficiency between the \emph{amorphous} and \emph{crystalline} cases increases with wavelength, between 0.23 (labeled on the figure) and 0.75 (obtained around $30^\circ$, for $\lambda=1200$ nm).To protect $\text{Sb}_2\text{S}_3$ from oxidation, the device is capped with a $\text{SiO}_2$ layer of thickness t (see Supplementary Information for details). Sweeping the capping thickness between 15 and 40 nm \cite{Feldmann:21,Cheng:23,Kong:23,Li:24} shows that the device performance is largely unaffected, with the passband and deflection efficiency decreasing by less than 2.1\% on average, highlighting the robustness of the design.

To further demonstrate the $\theta-\lambda$ range that our device enables in the presence of diffraction and to show the interplay of the diffraction modes, we present in Fig.~\ref{contours} the contour plot of the efficiency of both the $-1$st order and specular reflection over the $(\lambda,\theta)$-plane. The dashed black lines in the subplots are obtained using Eq.~(\ref{diff_rel}) and upper-bound regions for a diffraction order $n$ propagate. One can easily have a gauge of the loss mechanism present in the \emph{crystalline} nanorods from the low efficiency obtained in Fig.~\ref{contours} (c) vs (d) when only the 0-th order is reflected: in the first, there is total reflection while the latter presents region of almost no reflection (above $\lambda=2\:\mu$m.   

From Fig.~\ref{contours} (a) and (b), one notices that the efficiency $\eta_{(-1)}$ is non-zero only in regions where Eq.~(\ref{diff_rel}) dictates the propagation of the $-1$st order, which implies that the deflection process is limited by diffraction. Nevertheless, there is an increase in the passband with the incidence angle. Notably, the \emph{amorphous} case achieves significant efficiency only when the $-1$ order is the only non-zero order. In contrast, in the \emph{crystalline} case, despite the emergence of diffraction orders ($-2$nd and $+1$st), an efficiency of 50\% is still achieved, at short wavelengths (below $\lambda=1\:\mu$m).
Figure~\ref{contours} (a,c) establish a complementary domain where the power is dominantly reflected either as $-1$st order versus $0$th order, for the PCM is \emph{amorphous}. For its counterpart, the \emph{crystalline} case, the loss in the material limits the power of both the 0th order and the $-1$st order, while the former remains dominant. Since the 0th order or specular reflection is not nullified, even in the region where non-zero orders exist. This allows the device to be used as a power splitter as well.

\begin{figure}[t!]
 \centering
 \includegraphics[width=\linewidth]{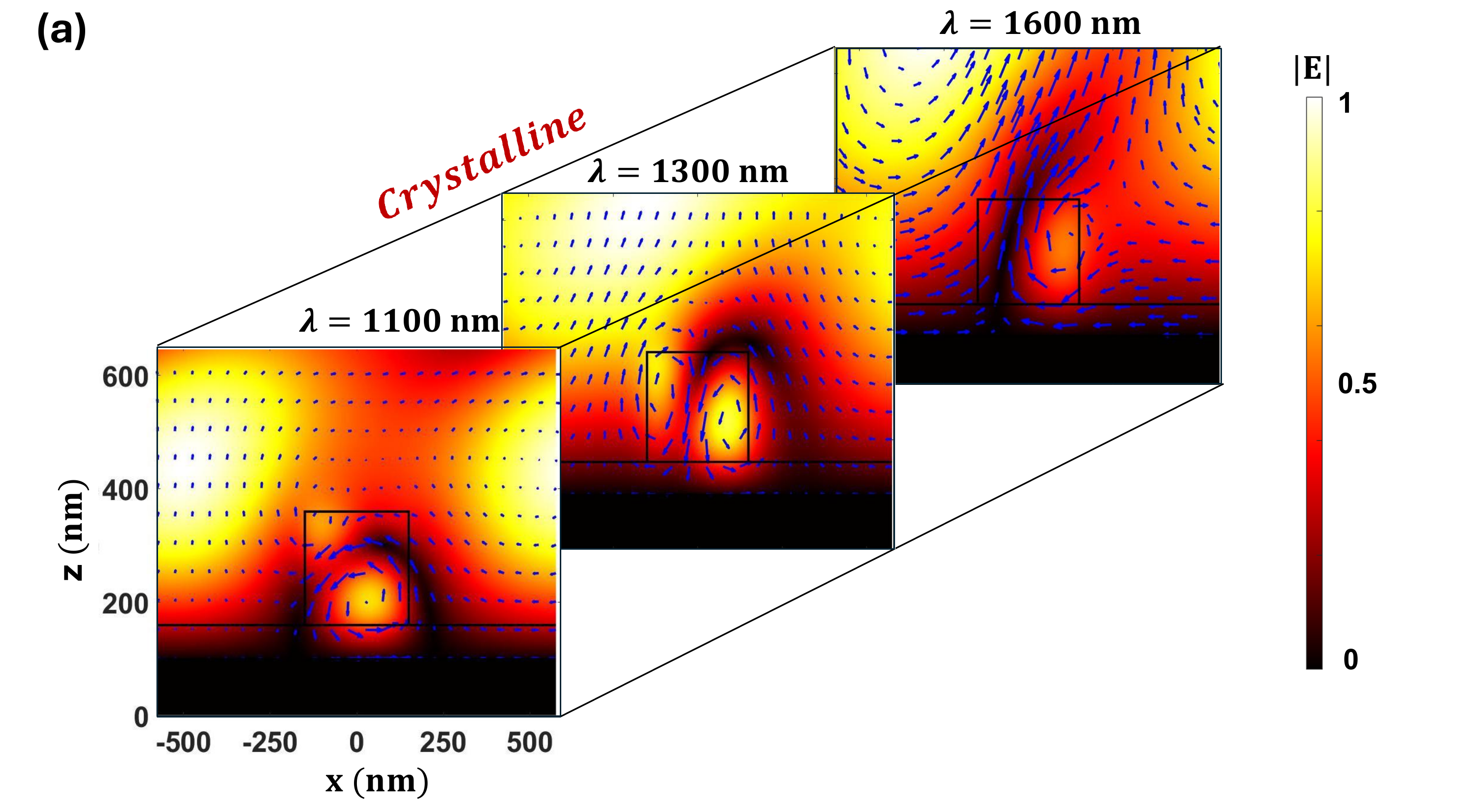}
 \includegraphics[width=\linewidth]{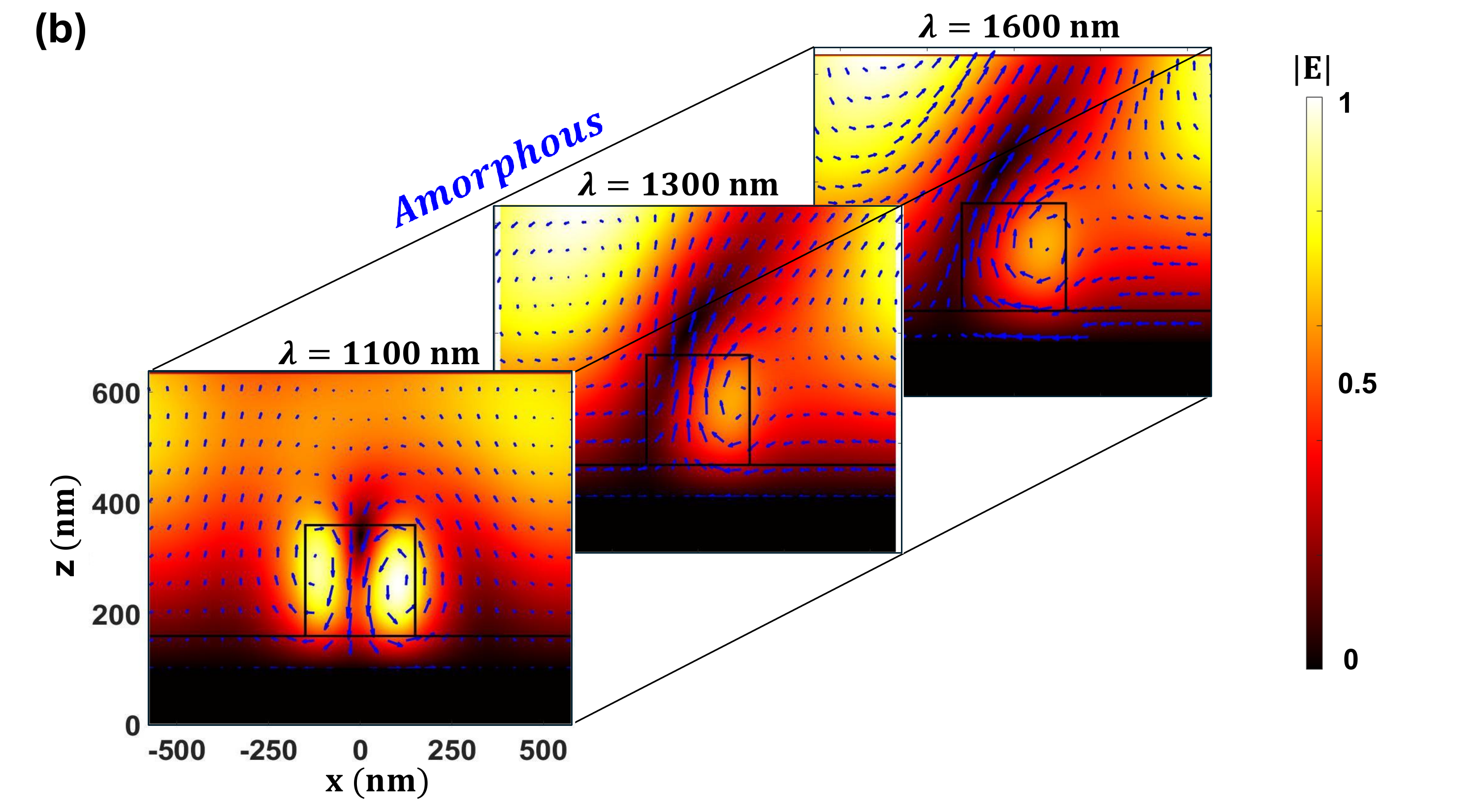}
\caption{ Electric field (colormap) and Magnetic field (quiver plot) inside and near the nanorod for \emph{crystalline} (a) and \emph{amorphous} (b) case at  $\lambda=1100 \rm \, nm$, $\lambda=1300 \rm \, nm$ and at $\lambda=1600 \rm \, nm$; the TE polarization is considered, for an incident angle $\theta=35^\circ$. The solid line (white) delimits the different layers of the structure.}
\label{Efields}
\end{figure}

To investigate the mode characteristics associated with the observed resonance, we plot the distribution of the electric field (E-field) under a TE incident wave, as shown in Fig.~\ref{Efields}.
The latter contrasts the E-field distribution around the nanorods in the \emph{crystalline} case (a) and the \emph{amorphous} case (b) for given wavelengths, and $\theta=35^\circ$. At $\lambda=1100$ nm, the E-field is nearly symmetric and localized in the nanorod in \emph{crystalline}  case (Fig.~\ref{Efields}(a)), as opposed to at its boundaries in \emph{amorphous} case (Fig.~\ref{Efields}(b)). In fact, in the \emph{crystalline} case, the E-field distribution reveals the mode pattern as of zero-order Mie resonance of a circular rod \cite{ZHAO2009,Valentine2017}. In contrast, in the \emph{amorphous} case, the distribution is consistent with a magnetic dipole resonance, as explicitly demonstrated by the magnetic field distribution in Fig.~\ref{Efields}(b). This is associated with a low efficiency (below 40\%) in both cases. Moreover, at $\lambda=$ 1300 nm, in both material phases (Fig.~\ref{Efields}), one notices a mode distribution similar to that of a tilted magnetic dipole with a weaker localization of the E-field inside the rod. 
The region of a weak electric field hosts a quasi-static magnetic field distribution that supports the excitation of a magnetic dipole.
This region being greater in the \emph{amorphous}-based structure yields, therefore, a larger equivalent magnetic dipole for greater efficiency. 
The tilted resonant magnetic dipole is also established at 1600 nm for both states. It shows a pronounced asymmetry in the field distribution, in agreement with the contrast between the H-field distribution at $\lambda =$1300 nm and 1600 nm (Fig.~\ref{Efields}(b)). We attribute the high efficiency of the -1st deflection coefficient to the asymmetry of the fields established in the device, namely, the excitation of a tilted magnetic dipole. In light of the significant similarities in the field distributions at 1300 nm and 1600 nm for the two states of the $\rm Sb_2S_3$, the difference in the $-1$st deflection efficiency can undoubtedly be attributed to the loss mechanism inside the \emph{crystalline} state.

\begin{figure}[h!]
    \centering
    \includegraphics[width=0.8\linewidth]{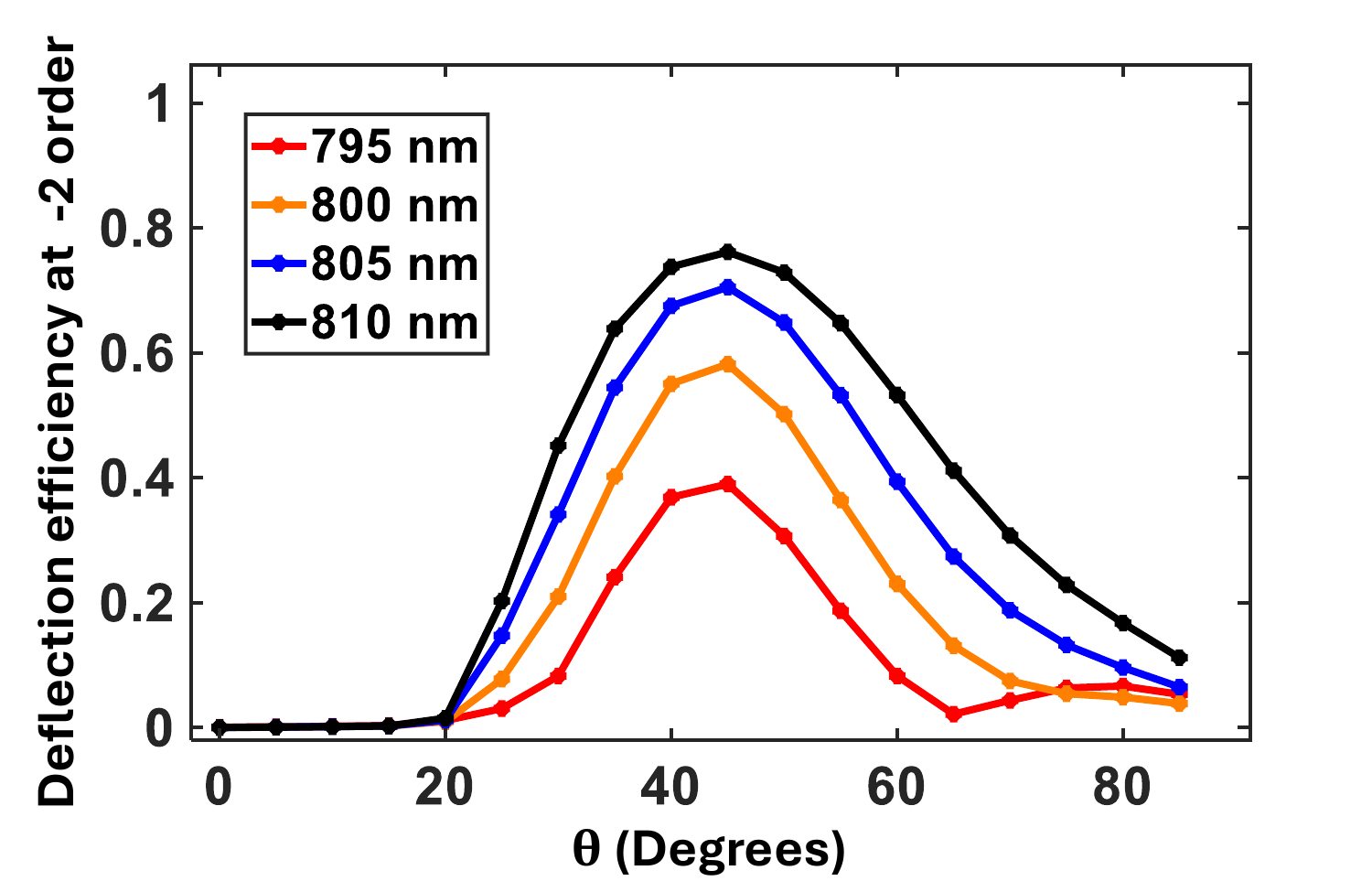}
    \caption{Narrow band pass spatial filtering exploiting -2nd order deflection in 795-810 nm, in TE polarization.}
    \label{neg_2_order_narrow}
\end{figure}

Further, the functionalities offered by our design go beyond wide-angle and ultrawide-band deflection and {\color{black} power attenuattion for high power application}. For the  \emph{amorphous}-based structure, the design ensures a spatial filtering, as hinted earlier, as an advantage of the sharp cutoff (Fig.~\ref{deflect_meas} (a,b)). From the existence of the bandpass with sharp cutoffs, our design demonstrates a frequency selectivity that characterizes spatial filtering \cite{spatialfiltering}. Through spatial filtering, one can sort the incident waves and, as a result, modify the angular spectrum \cite{Peri1985}. While the -1st order provides reflection mainly at large angle $\theta$ (Fig.~\ref{deflect_meas} b), the -2nd order deflection from the nanorod structure results in a narrow band-pass-spatial filtering. Figure~\ref{neg_2_order_narrow} shows that for a TE polarization,  an angular bandwidth of 40$ ^\circ$ - 60$ ^\circ$ is obtained around $\lambda=800$ nm.


\section{Conclusion}
In summary, we have numerically demonstrated a highly efficient and multifunctional reflective meta-array based on $\rm Sb_2S_3$ phase-change material nanorods operating in the near-infrared regime. The proposed design achieves ultra-wideband and wide-angle deflection in the \emph{amorphous} state, with deflection efficiencies exceeding 80\% over more than 1000 nm of bandwidth and incidence angles spanning up to $55^\circ$. While in the \emph{crystalline} state, the inherent material loss enables effective optical power limiting, reducing the deflected power below 60\%, with values as low as 20\% in the O-band. Additionally, the device exhibits functionalities such as spectral filtering, wide-angle deflection, and partial power splitting, with complementary deflection and reflection characteristics.

The field distribution analysis further highlights the role of low-order Mie resonances and their evolution with wavelength and phase state, providing insights into the underlying physical mechanisms. Thanks to its simplicity, robust performance, and dynamic tunability, the proposed meta-array offers a promising platform for applications in optical power limiting, beam shaping, and integrated photonic systems, particularly for applications requiring broadband operation and high power resilience.

\begin{backmatter}

\bmsection{Disclosures} The authors declare no conflicts of interest.

\bmsection{Data availability} Data underlying the results presented in this paper are not publicly available at this time but may be obtained from the authors upon reasonable request.
\end{backmatter}


\bibliography{main}

\bibliographyfullrefs{main}


\end{document}